# Enhanced strength–ductility combination by introducing bimodal grains structures in high-density oxide dispersion strengthened FeCrAl alloys fabricated by spark plasma sintering technology

Xu Yan[a], Zhifeng Li[a], Haoxian Yang[b,*]，Sheng Wang[a,c,*]

[a] School of Nuclear Science and Technology, Xi'an Jiaotong University, Xi'an, 710049, Shaanxi, China.

[b] Guangdong Province Hospital for Occupational Disease Prevention and Treatment, Guangzhou, 510300, Guangdong, China.

[c] XJTU-Huzhou Neutron Science Laboratory, Science Valley Medium-sized Building, Huzhou, 313000, Zhejiang, China.

*Corresponding author: Haoxian Yang, E-mail: gdfsyhx@163.com;

Sheng Wang, E-mail: shengwang@xjtu.edu.cn.

**Abstract**

Oxide dispersion strengthened (ODS) FeCrAl alloys dispersed high-density nano-oxides in the matrix show outstanding corrosion resistance and mechanical properties. However, ODS FeCrAl alloys achieve the high strength generally at the expense of ductility in some way. Here, a method by introducing a bimodal grain structure was designed to overcome the strength-ductility tradeoff. In this work, ODS FeCrAl alloys were successfully fabricated through various mechanical alloying (MA) time, combined with spark plasma sintering (SPS) under the vacuum of less than 4Pa. Microstructural characterization showed that the average grains size and nano-oxides size decrease gradually, and the density of nano-oxides increases, as the milling time increases. Mechanical properties revealed that both the strength and ductility were significantly synergistic enhanced with increasing milling time. The bimodal grain distribution characterized by electron backscatter diffraction (EBSD) (vacuum degree was less than $5\times10^{-5}$pa) was beneficial for the activation of the back stress strengthening and the annihilation of these microcracks, thus achieving the excellent ductility (27.65%). In addition, transmission electron microscope (TEM) characterization under the vacuum degree of less than $10^{-6}$pa illustrated that ultra-high-density nano-oxides ($9.61\times10^{22}/m^3$) was crucial for enhancing the strength of ODS FeCrAl alloys (993MPa). The strengthening mechanism superposition, based on the model of nano-oxides interrelated with the dislocation, illustrated an excellent agreement with experimental results from yield strength strengthening mechanisms. To our best knowledge, H40 (milled for 40h, and sintered at 1100 °C) alloy presents the outstanding strength with the exceptional ductility among all studied ODS FeCrAl alloys, which makes it the promising cladding materials for the accident tolerant fuel (ATF) cladding.

## 1. Introduction

Since the devasted Fukushima Daiichi nuclear reactor accident in 2011, great attention has been focused on the developing the new and safe accident-tolerant fuel (ATF) cladding materials [1]. The ATF materials should possess the outstanding mechanical properties and excellent corrosion resistance at high temperatures, which will ensure the fuel integrity to prolong the service periods of severe scenarios, thus reduce the loss and increase the reactor safety [2].

Oxide dispersion strengthened (ODS) steels dispersed the high-density nano-oxides in the matrix are considered to be one of the prospective candidates cladding materials for nuclear applications [3]. These high-density nano-oxides and nano-clusters (NCs) are homogenously distributed in the matrix, serving as the pinning points to prevent grain boundary migration and dislocation motion, which will contribute to enhancing the strength [4]. Up to now, ODS Fe-Cr alloys are the most widely investigated ODS steels, and they are also effective for corrosion resistance, due to the $Cr_2O_3$ formed on the surface [5]. However, the oxide layer of $Cr_2O_3$ is unstable over 1000 °C, which greatly limits the application in high temperatures. More seriously, ODS steels with high Cr content are likely to induce Cr-rich α' phase at 300 °C-350 °C in light water reactors, resulting in thermal aging embrittlement, which will be ductility reduction and a poor fracture toughness [6]. To address this problem, many scholars investigated and observed that the addition of Al element in ODS Fe-Cr alloys will significantly improve the corrosion resistance up to 1400 °C to form the dense $Al_2O_3$ protective layers [7]. Nevertheless, much more coarsening nano-oxides are dispersed in ODS FeCrAl alloys with a lower number density than that of ODS FeCr alloys, since the addition of Al in ODS FeCr alloys will shift nano-oxide types from high stable Y-Ti-O phase to large-size Y-Al-O phase, which will show less resistance to coarsening and decrease strength in the high temperature environments [8]. Many efforts have made to identify that Zr element in ODS FeCrAl alloys will change the types of nano-oxides from coarse Y-Al-O phase to fine Y-Zr-O phase, since the binding energy of Y-Al-O is lower than that of Y-Zr-O. This means that Y-Zr-O is usually easier to form and more stable than Y-Al-O [9]. Thus, the addition of Zr element in ODS FeCrAl alloys could refine the nano-oxides. Above all, ODS FeCrAlZr alloys possess good mechanical properties and strong corrosion resistance in the high temperatures, making it a promising candidate material for ATF materials [9].

However, how to further enhance the mechanical properties in ODS FeCrAl alloys is still being the urgent and eager pursuit, when serviced in a severe reactor operating environments. In the last few decades, many researchers have been made efforts to enhance the strength and ductility of ODS FeCrAl alloys. The preparation technology of mechanical alloying (MA) and hot extrusion (HE) were employed to produce Fe-12Cr-5Al ODS alloys from the Oak Ridge National Laboratory. The ultimate stress was up to 1300MPa while the elongation was only about 10%, and it was depicted that the improvement in the tensile strength was attributed to the ultra-high-density nano-oxides ($10^{23}/m^3$) [10]. In addition, Li, et, al. studied the effect of hot rolling and annealing temperatures on mechanical properties in Fe-15Cr-4.5Al ODS alloys and the results displayed the ultimate stress was increasing to about 1200MPa yet the uniform elongation was less than 13%. It was stated that the improvement of strength was ascribed to the small nano-oxides and fine grain sizes [11]. Zhou, et, al. added the hot forging technology after MA and HE in Fe-13Cr-5Al ODS alloys to improve the mechanical properties. The elongation was about 20%, while the ultimate stress was decreased to be less than 700MPa, and the ultimate stress was obviously increased to be more than 1000MPa with a lower elongation of 16%, when 0.6% Zr element was added in the matrix [12]. Meanwhile, Ukai, et, al. studied the effect of the content of Cr and Al elements on the mechanical properties for ODS FeCrAl alloys, since both Cr and Al were vital elements with synergistic effects in strength and ductility. The results illustrated that Fe-14Cr-12Al ODS alloys exhibited an outstanding ultimate stress of about 1100MPa with the elongation less than 20% [13].

To sum up, there usually existed a trade-off between the strength and ductility of ODS FeCrAl alloys. However, to improve the strength without sacrificing the ductility and even further promote the ductility are always being a quite challenge. Therefore, in this study, we employed an advanced Spark Plasma Sintering (SPS) consolidated technology under the vacuum of less than 4Pa to fabricate the ODS FeCrAl alloys and to probe into the mechanical behavior under the various milling times. Multiply characterization methods suggest that the high-density nano-oxide with small sizes characterized by TEM (the vacuum degree was less than $10^{-6}$Pa), and the bimodal grain distribution characterized by EBSD (the vacuum degree was less than $5\times10^{-5}$Pa) enormously enhance the strength and ductility of ODS FeCrAl alloys, which will provide value experience for the subsequent preparation of outstanding strength and excellent ductility in nuclear materials.

## 2. Experiment
### 2.1. Materials and preparation

The gas-atomized pre-alloyed powders of FeCrAlZrTi with the particle size between 40 μm and 150 μm were mechanical alloyed with 0.3% (wt.%) $Y_2O_3$ powders under the argon atmosphere in a planetary high-energy ball mill using 304 stainless steel balls (10 mm diameters and 5 mm diameters) in 304 stainless steel milling media. The powders were milled at a speed of 500 rpm for 5min–80h with the ball to powder ratio of 10:1. These milling conditions were further selected based on the powder morphology and size on MA. The milled powders were sintered by SPS (LABOX-325R, Japan) at 1100 °C for 7 min under the pressure of 50 MPa with the vacuum of less than 4Pa, and then cooled to room temperatures with a cooling rate of 100 °C/min.

The bulk specimens were prepared by wire-cut electric spark with the surface of 7 mm × 7 mm and the thickness of 1.5 mm. The surface was then mechanically ground up to 4000 grids with SiC papers, and then was polished by alumina paste. A vibratory polishing and ion thinning at room temperatures were also employed on polished specimens to eliminate the stress from samples surface. In addition, samples from transmission electron microscope (TEM) were prepared by electrochemical thinning in twin-jet polisher using 10% perchloric acid and 90% alcohol solution under the voltage of 20V at −21 °C.

## 2.2. Microstructure characterization

The density of the sintered samples was measured at room temperatures by the Archimedes method for ten times to reduce the experiment errors, and the microstructure were also examined by optical microscopy (OM). X-Ray diffraction (XRD) was conducted on a Bruker D8 Advanced diffractometer to identify constituent phases, and a step size of 0.02° with a measuring time of 0.1 s per step was employed under a current of 40 mA and a voltage of 40 kV. The morphology of the powders and fracture surfaces were observed in Gemini SEM 500 field emission scanning electron microscopy (SEM). In addition, the grain morphology was characterized by electron backscatter diffraction (EBSD) equipped with JEOL 6500F SEM with step length of 180 nm under the vacuum of less than $5×10^{-5}$pa. The density and size of nano-oxides were characterized by the Talos-F200 TEM at 200 kV under the vacuum of less than $10^{-6}$pa. The crystal structure of nano-oxides was characterized by high resolution transmission electron microscopy (HRTEM). The energy dispersive spectroscopy (EDS) with the STEM mode with a camera length of 125 mm and the spot size of 5 was conducted to characterize the composition of nano-oxides.

## 2.3. Mechanical property test

Tensile property test was conducted at room temperatures using an MTS hydraulic tensile machine at a constant strain rate of $10^{-3}$/s. The size of the tensile sample was shown in Fig. 1. A minimum of five samples were performed to test and verify the results.

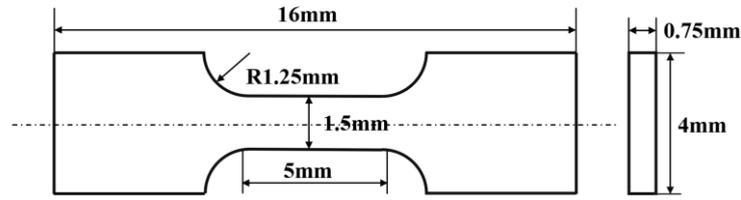

**Fig. 1.** The geometry of the tensile specimen.

## 3. Result
### 3.1. XRD and SEM characterization of milled powders

Fig. 2 illustrated the XRD results of powders (FeCrAlZrTi+$Y_2O_3$) at various milling times. After 5 minutes of MA, Fe diffraction peak was obvious, and $Y_2O_3$ peak was weak. The diffraction peak of $Y_2O_3$ disappeared up to 30 minutes, indicating that $Y_2O_3$ gradually dissolved into the matrix. There only existed Fe diffraction peak after 30 minutes in each curve, meaning that there was no detected other phase structure in the milling process. With the increase of milling time, the width of diffraction peak increased gradually, which was mainly ascribed to grain refinement and lattice distortion [14]. Since the width of the diffraction peak was inversely proportional to the grain size and directly proportional to the lattice distortion, this indicated that the grain size of powder decreased and the lattice distortion increased gradually with the milling time.

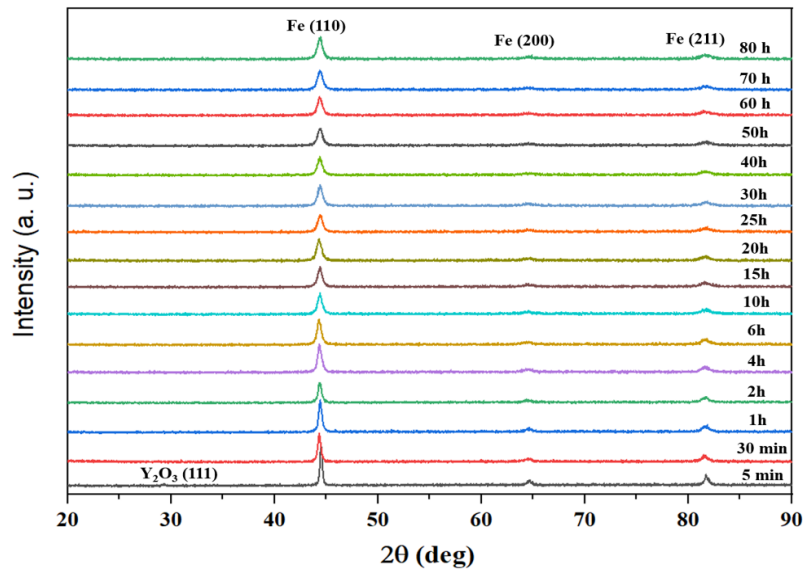

**Fig. 2.** XRD patterns of the as-milled powders at various milling time.

The SEM morphologies of powders at various milling times were shown in Fig. 3, which distinctively exhibited the progressive changes in morphology. Fig. 4 collected the variation of particle sizes with the milling time. Powder particles underwent repeated flattened and cold welding during 5 mins to 10 h and the mean size reached to the maximum of 120 μm at 10h milling. During the milling period of 10h-40h, powder particles were exposed to repeated severe fragmentation, and the large-size powders were milled into small-size powders, and the average size of the powders decreased to 35 μm in 40h. Continuing to increasing the milling time from 40h to 80h, the size of the powder almost no changed, and this process had reached a balance of powder fragmentation and cold welding.

As we all know, MA is one of the key steps for the fabrication in ODS steels. The effect of different MA times on the powder size, morphology and subsequently on the distribution of grain size and nano-oxides in consolidated samples is critical, which in turn directly affects the microstructure and mechanical properties [15]. Combined with the Fig. 3 and Fig. 4, we have identified as followed. Both the mean size of powders and the difference among particles were maximized at 10 h in cold welding stage. In addition, in the crushing stage, the reduction rate of powder size was significantly reduced after 20h. Moreover, the powder size at 40h in the balance stage basically remained unchanged. It would introduce serious powder contamination from the milling media more than 40h milling. Therefore, the powders milled at 10h, 20h, and 40h were selected to further be sintered at 1100 °C by SPS. For the convenience of distinguishing various types of ODS FeCrAl alloys, 10H signified the sample which was milled for 10h and then sintered at 1100 °C, 20H indicated the sample which was milled for 20h and then sintered at 1100 °C, and 40H expressed the sample which was milled for 40h and then sintered at 1100 °C.

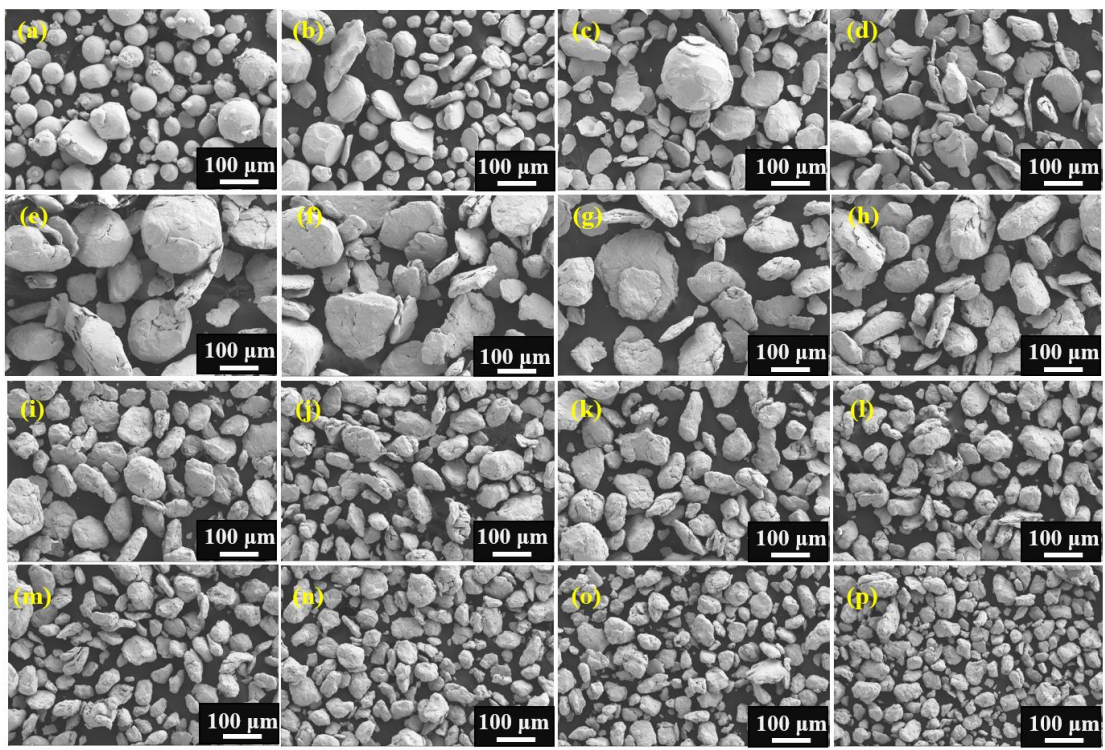

**Fig. 3.** The SEM images of as-milled powder (a) 5min, (b) 30min, (c) 1h, (d) 2h, (e) 4h, (f) 6h, (g) 10h, (h) 15h, (i) 20h, (j) 25h, (k) 30h, (l) 40h, (m) 50h, (n) 60h, (o) 70h, (p) 80h.

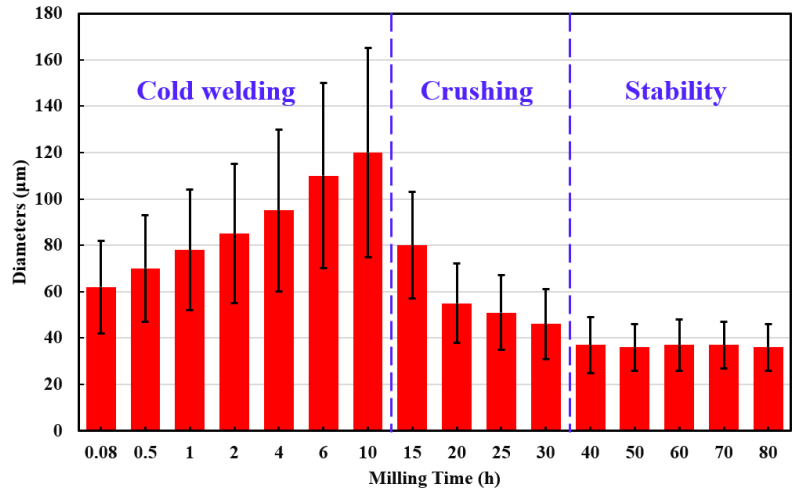

**Fig. 4.** The particle sizes variation as a function of the milling time.

### 3.2. Optical microscopy characterization of sintered samples

The optical micrographs of sintered samples were shown in Fig. 5. There were only little small pores rather than large-size pores in the matrix, which meant that the sintering performance was dense and efficient. During SPS, the discharge between particles generated high temperature, inducing the diffusion and migration of solute atoms, and then filling the intergranular pores [16]. In addition, relative densities of samples of 10H, 20H, and 40H in Tab. 1 were 95.45%, 96.05%, and 95.66%, respectively, by the method of Archimedes drainage method. The relative density of the samples sintered at 1100 °C was all exceed 95%, which was coincidence to the results of OM characterization.

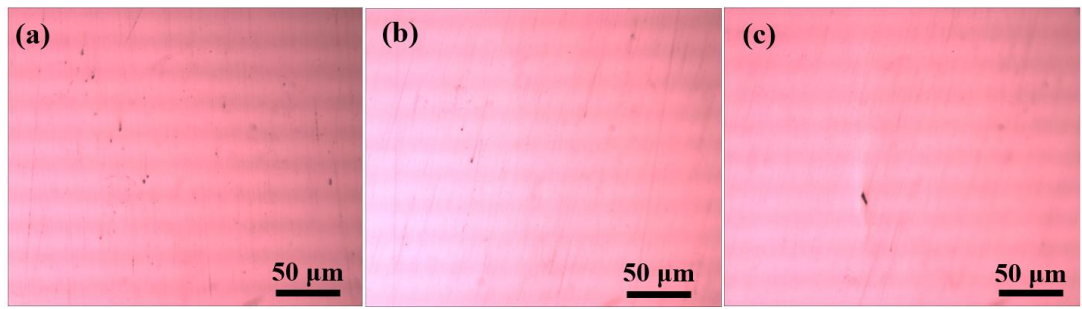

**Fig. 5.** Optical micrographs of ODS FeCrAl alloys (a) 10H, (b) 20H, (c) 40H. (10H, 20H, and 40H represented that the samples were milled 10h, sintered at 1100 °C; milled 20h, sintered at 1100 °C; and milled 40h, sintered at 1100 °C, respectively.)

**Tab. 1.** The relative density for different ODS FeCrAl alloys.

| Samples | 10H | 20H | 40H |
|---|---|---|---|
| **Relative Density (%)** | 95.45±0.06 | 96.06±0.09 | 95.65±0.09 |

### 3.3. XRD characterization of sintered samples

Fig. 6 showed the XRD patterns for various ODS FeCrAl alloys. The result showed that the prepared ODS FeCrAl alloys were all single Bcc-α-Fe structures. In addition, the phase of nano-oxides was not detected due to the influence of XRD resolution.

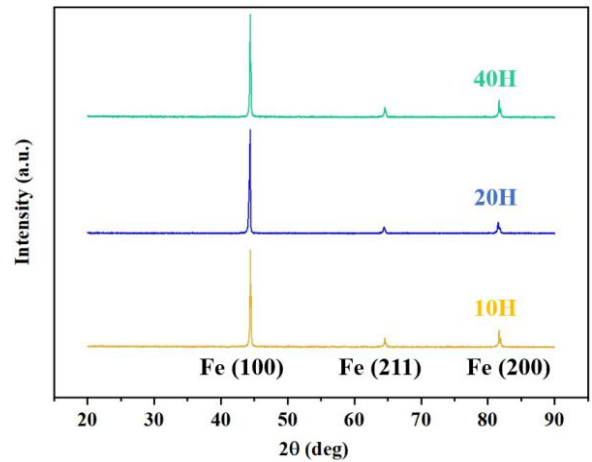

**Fig. 6.** The XRD patterns for different ODS FeCrAl alloys.

### 3.4. EBSD characterization

General microstructure characterized by EBSD consisted of a bimodal grain size distribution with a ferritic structure, as shown in Fig. 7. During SPS, the inhomogeneous temperature gradient contributed to the bimodal grain distribution [16], so plenty of ultrafine grains and some coarse grains in this study were observed in Fig.7 (a-c). In addition, the EBSD results have showed that the grain sizes decreased with increasing the milling time, and the average grain sizes in Fig.7 (d-f) were 15μm, 14.3μm, and 10.2μm for 10H, 20H, and 40H, respectively. More interestingly, the size difference of grain would reduce gradually with the increase of the milling time.

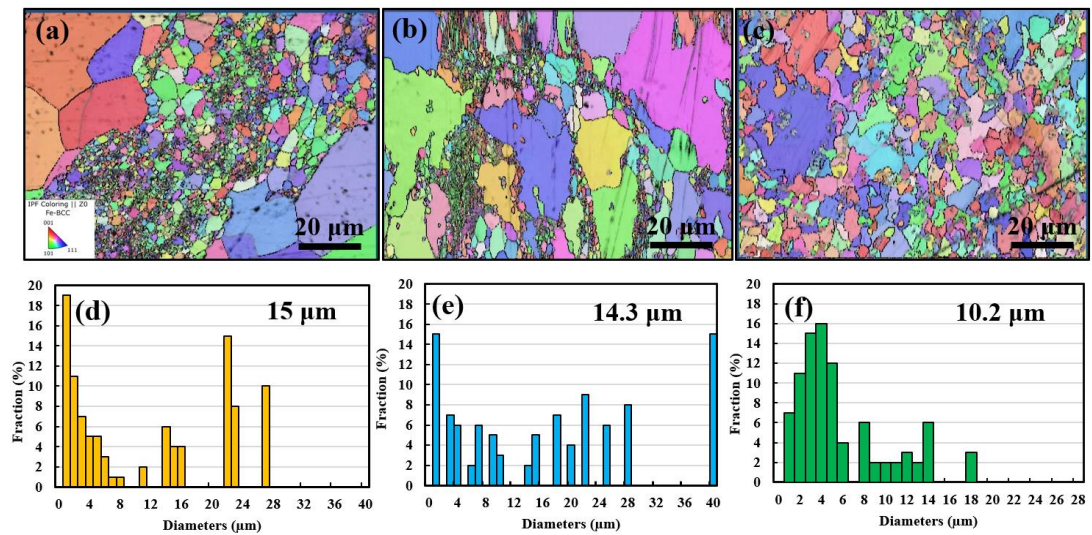

**Fig. 7.** (a-c) IPF images from EBSD characterization for 10H, 20H, and 40H,

respectively, (d-f) the distribution of grain sizes for 10H, 20H, and 40H, respectively.

## 3.5. TEM/STEM characterization

The typical bright field (BF)-TEM images of ODS FeCrAl alloys were shown in Fig. 8. It could be seen that some nano-oxides of various sizes were dispersed in the ferritic matrix, and other nano-oxides pinned the grain boundaries and dislocation lines, which would greatly improve the mechanical strength in ODS steels. High angle annular dark field (HAADF) images of nano-oxides in ODS FeCrAl alloys were characterized in Fig. 9. High-density nano-oxides with various size were distributed in the matrix. The size and density distribution of the nano-oxides was depicted in Fig. 10. The average diameters of nano-oxide were 27.2 nm in 10H, 13.1nm in 20H, and 5.9 nm in 40H, respectively. The density of the nano-oxides were $1.44\times10^{22}/m^3$ in 10H, $5.70\times10^{22}/m^3$ in 20H, and $9.61\times10^{22}/m^3$ in 40H, respectively. Apparently, the size of the nano-oxides decreased and the density increased with increasing the milling time.

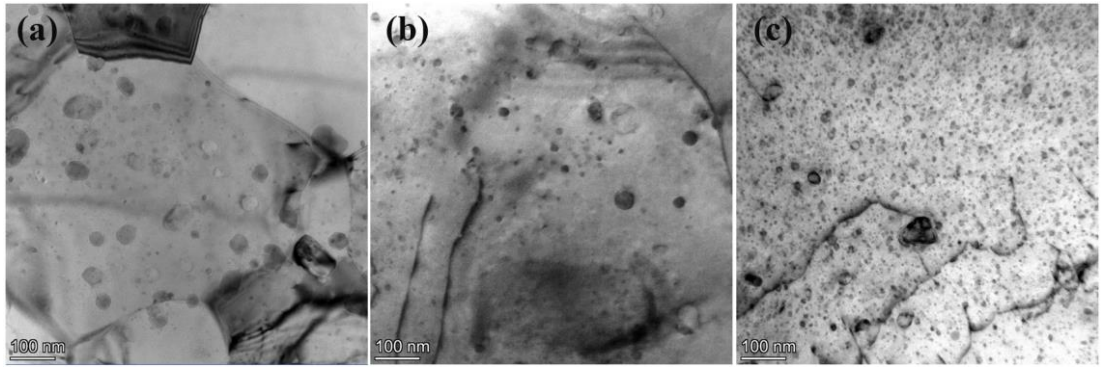

**Fig. 8.** BF-TEM images (a) 10H, (b) 20H, and (c) 40H.

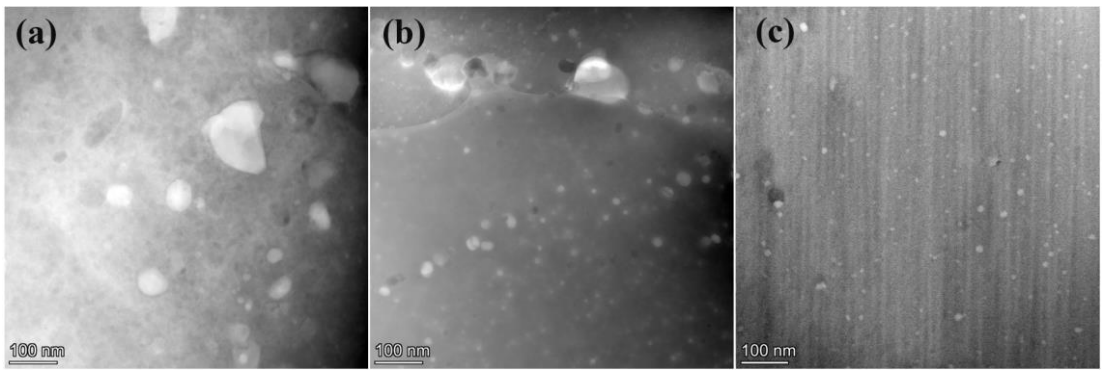

**Fig. 9.** HAADF-STEM images (a) 10H, (b) 20H, and (c) 40H.

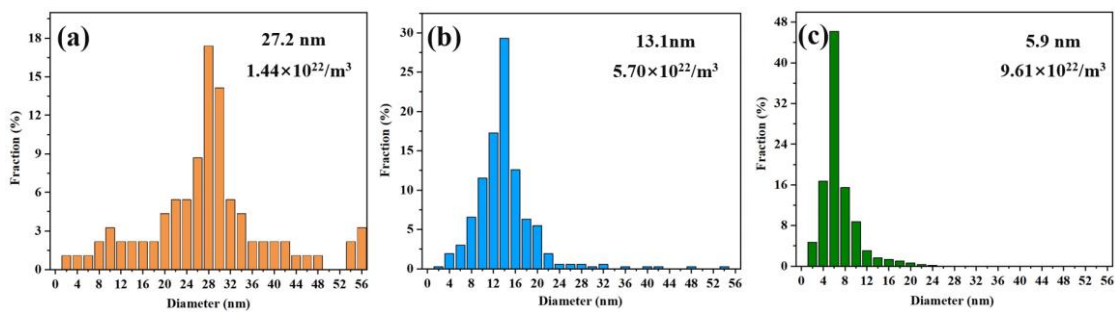

**Fig. 10.** Size distributions of the nano-oxides in ODS FeCrAl alloys (a) 10H, (b) 20H, and (c) 40H.

HAADF images in 20H illustrated that the nano-oxides were homogenously distributed in the matrix as revealed in Fig. 11(a). The EDS mappings in Fig. 11(b) to 11(h) depicted that Al, O, Ti, Y, and Zr were enriched, while Cr and Fe were depleted in the nano-oxides. Careful observation found that the large-size nano-oxides were more likely to form Y-Al-O, while the small-size nano-oxides were prone to form Y-Ti-O and Y-Zr-O. And the results were consistent with the literature [17]. Crystalline structures from the nano-oxides were confirmed by HRTEM in Fig.12 (a-d) and corresponding fast fourier transformations (FFT) in Fig.12 (e-h), respectively. The results identified that there were four types of nano-oxides, which were $Y_3Al_5O_{12}$, $Y_2Zr_2O_7$, $YTiO_3$, and $Y_4Zr_3O_{12}$.

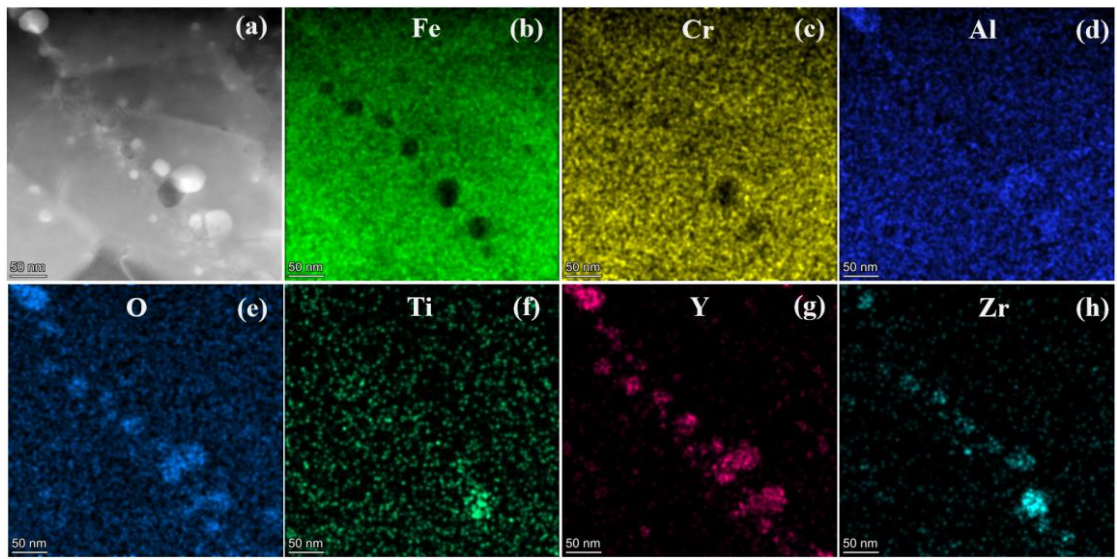

**Fig. 11.** (a) HAADF images of nano-oxides in 20H, (b–h) Fe, Cr, Al, O, Ti, Y, and Zr EDS mappings, respectively.

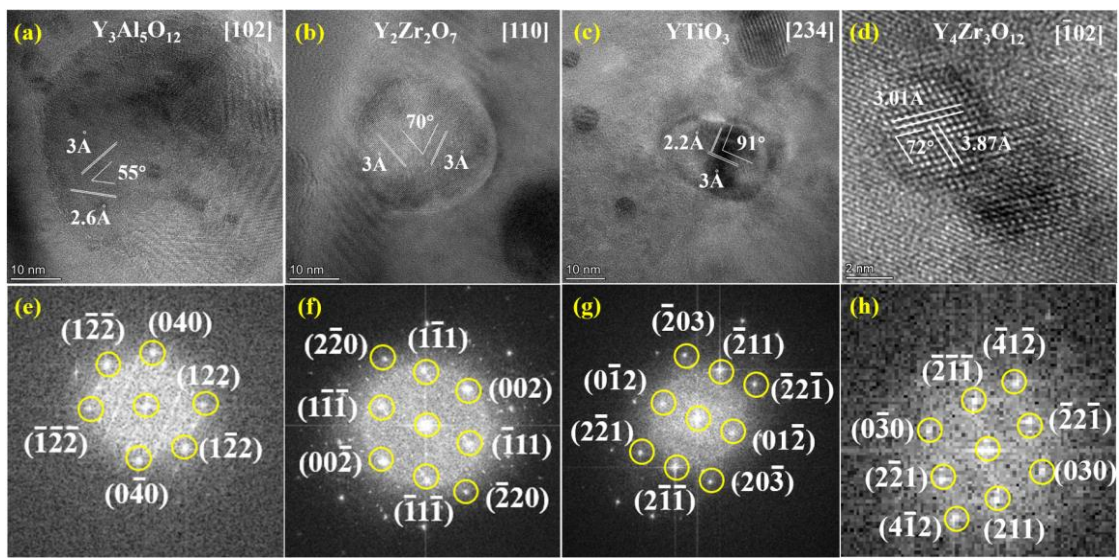

**Fig. 12.** HRTEM images and corresponding FFT diagrams of nano-oxides (a, e) $Y_3Al_5O_{12}$, (b, f) $Y_2Zr_2O_7$, (c, g) $YTiO_3$, (d, h) $Y_4Zr_3O_{12}$.

### 3.6. Tensile test

Fig. 13 depicted the stress-strain curves of three kinds of the ODS FeCrAl alloys at room temperatures. More than five tensile tests were conducted on each sample. The tensile properties of ultimate stress (UTS), yield stress (YS), and total elongation (TE) were summarized in Tab. 2. It showed that UTS, YS, and TE in 10H were identified as 853MPa, 702MPa, and 18.54%, respectively. UTS, YS, and TE in 20H were revealed as 916MPa, 789MPa, and 24.30%, respectively. UTS, YS, and TE in 40H were shown as 993MPa, 872MPa, and 27.65%, respectively. To our surprise, both the strength and ductility of the ODS FeCrAl alloys increased with increasing the milling time, which signified that the ductility did not decrease but increase, as the alloy strength increased. More details on the strength and ductility were further discussed in the Discussion section.

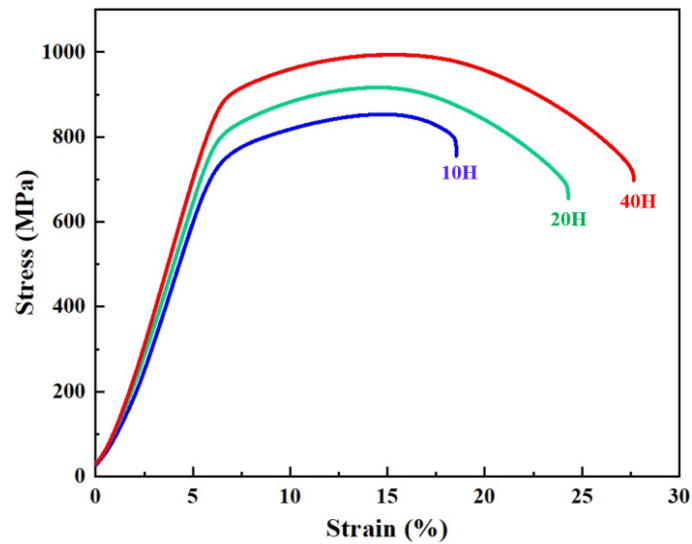

**Fig. 13.** Strain-stress curves of various ODS FeCrAl alloys at room temperature.

**Tab. 2.** The results of tensile tests for various ODS FeCrAl alloys.

| Samples | UTS (MPa) | YS (MPa) | TE (%) |
| --- | --- | --- | --- |
| 10H | 853 | 702 | 18.54 |
| 20H | 916 | 789 | 24.30 |
| 40H | 993 | 872 | 27.65 |

## 4. Discussion
### 4.1. Grains

The grain sizes in this study revealed a typical bimodal distribution. During the sintering process, the current more easily passed along the surfaces of the particles than that of the cores because of a lower electric resistance, so the temperature distribution in particles was not uniform. The temperature along powder surface was high and the grains were corresponding coarse. Inversely, the current density was low at the cores of the powders, and the grains cannot grow quickly enough, resulting in a small grain [17].

In addition, the deformation degree of various particles was uneven during milling, and the recrystallization could lead to rapid grain growth at the expense of the high energy and small size grains. Moreover, the mean particle size in 40H was small and range of the particle size was narrow, which contributed to a relatively uniform recrystallization process [18]. Thus, 40H, compared to 20H and 10H, possessed a smaller difference in bimodal grain size.

Further, the oxide formers elements could combine with the oxygen during SPS to form the small nano-clusters and the nano-oxides, which would pin the grain boundaries to prevent the grain growth. However, the difference in the uneven alloying degree would result in the different precipitation of nano-oxides. The grain size distribution was more homogeneous due to the stronger pinning effect, when the nano-oxides were more uniform at a longer milling time [19].

### 4.2. Oxides

The size of the nano-oxides decreased and the number density increased in ODS FeCrAl alloys, when the milling time increased from 10h to 40h. Actually, $Y_2O_3$ were gradually dissolved into the alloy matrix during the MA process, and the size of powders would be further refined and much more vacancies and dislocations would be occurred, as milling progressed [20]. Subsequently, more dislocations and vacancies could provide more abundant nucleation sites for nano-oxides and nano-clusters to precipitate during SPS due to the higher stored energy and thermal effect [21]. Consequently, smaller and denser nano-oxides in 40H would precipitate in the matrix than that of 10H and 20H. In reverse, these finer-size nano-oxides could also act as more robust pinning effects that stabilize the grains by impeding the grain boundary movement than that of coarsen nano-oxides, realizing smaller-size grains in 40H, which was coincidence to the result of EBSD characterization from Fig. 7.

The results of EDS with HRTEM characterization clarified that the large nano-oxides were Y-Al-O, while the small nano-oxides were Y-Zr-O and Y-Ti-O. This was agreement with the results from Rahmanifard's study, and they have determined that nano-oxides less than 10nm were the Y-Ti-Zr-O [22]. Meanwhile, Zhou, et, al investigated the effect of Al on the structures of nano-oxides in14Cr ODS alloys, and also found that Y-Al-O was the large nano-oxides [23]. In fact, various solute atoms possessed different affinity with oxygen, resulting in various types of nano-oxides. The elements of Zr and O had higher affinity than that of Al and O, Y-Zr-O was therefore easier to generate than Y-Al-O [24]. In addition, the bond energy of cluster in Y-Zr-O (0.85 eV) was higher than of Y-Al-O (0.1eV) and Y-Ti-O, so the nano-oxides of Y-Zr-O were prone to occur [9]. The formations of $Y_4Zr_3O_{12}$, $Y_2Zr_2O_7$ revealed the internal oxidation reaction as follows: $2Y_2O_3+3ZrO_2 \rightarrow Y_4Zr_3O_{12}$, $Y_2O_3+2ZrO_2 \rightarrow Y_2Zr_2O_7$. Dou, et, al also stated that both $Y_4Zr_3O_{12}$ and $Y_2Zr_2O_7$ were the common types of Y-Zr-O, when they surveyed the element of Zr and Ti addition on the structures of nano-oxides in ODS FeCrAl alloys [17]. Meanwhile, we have observed that there remained some nano-oxides of Y-Ti-O ($YTiO_3$). However, the addition of Al would inhibit the combination of elements of Ti and O, and preferentially precipitated Y-Al-O, since the formation energy of Ti (-77KJ/mol) oxide was lower than that of Al (-80 KJ/mol) [25]. In addition, the addition of Ti will delay the coarsening of Y-Al-O dispersion phase, and some small-size Y-Ti-O could also occur [26]. Al element in atomized powder could react with O element during ball milling: $2[Al]+3[O] \rightarrow Al_2O_3$, $Al_2O_3$ reacts with $Y_2O_3$ in the subsequent sintering process as follows: $5Al_2O_3+ 3Y_2O_3 \rightarrow 2Y_3Al_5O_{12}$. Nevertheless, the structure of Y-Al-O is inconsistent with the literature, which followed as $Al_2O_3+ 2Y_2O_3 \rightarrow Y_4Al_2O_9$ [27]. This difference might be due to the various preparation parameters (especially in terms of differences in consolidated temperature) in these two kinds of ODS FeCrAl alloys.

**4.3. Ductility**

The fracture surface was further characterized by SEM to understand the failure patterns from tensile tests. Fig. 14(a), (e), and (i) presented the overall surface fracture morphologies, and Fig. 14(b-d), (f-h), and (j-l) depicted the local microscopic morphology properties. Fig. 14(b-d) showed that the fracture surface possessed the cleavage area and ductile area, which implied that fracture mechanism of 10H was the coexistence of brittle and ductile fracture mode. However, the proportion of the ductile area in Fig. 14(f) would be greatly increased in 20H. And in Fig. 14(e), it could be observed that there existed river-like layers in some ductile regions. In addition, lots of the dimples from tens of micrometers to hundreds of nanometers were occurred in Fig. 14(g, h), which was regarded as the dimple ductile fracture mode. In Fig. 14(i), it could be seen that there were much more ductile regions from various layers in 40H than those in 20H, which meant that cracks were more likely to pass through various layers and showed a river-like fracture during tensile tests. In addition, the transfer of cracks between different layers required high energy, leaving a large number of dimples [12]. Thus, high density dimples appeared in Fig. 14(k), and the size of the dimples was uniform of tens of nanometers in Fig. 14(l), which was significantly smaller than those of 20H. Therefore, the fracture mechanism in 40H was the typical dimple ductile fracture pattern with an excellent plasticity.

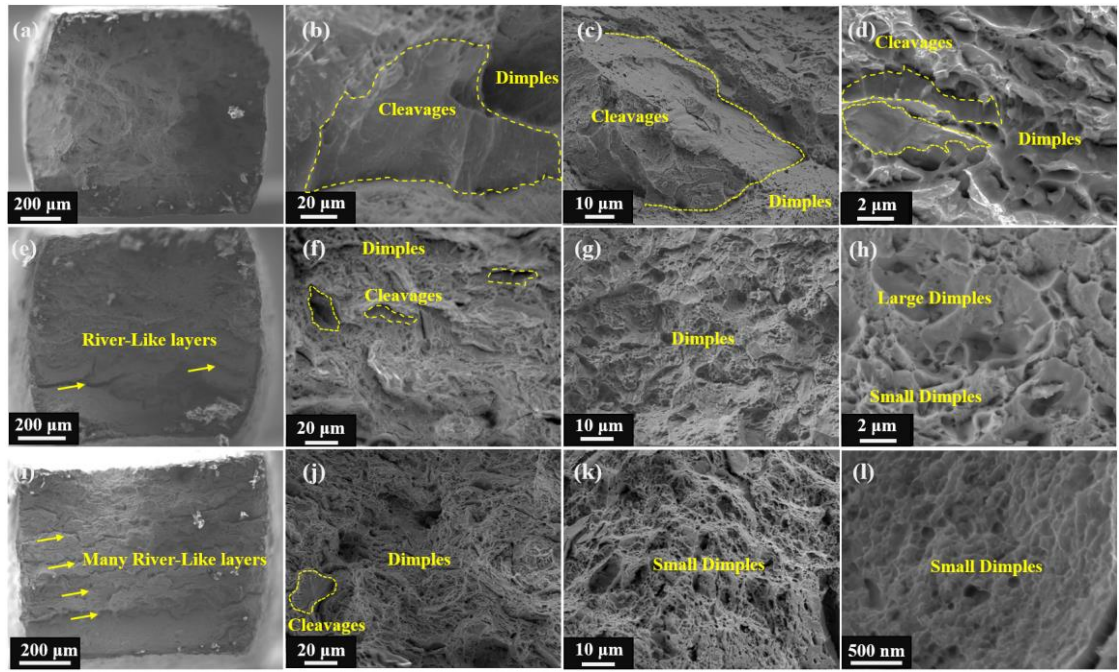

**Fig. 14**. SEM images of fracture surfaces of the various specimens after tensile tests (a-d) 10H, (e-h) 20H, (i-l) 40H.

Bimodal grains in this work were also crucial significance for the improving the strength and ductility. The regions between the fine grains and the coarse grains existed the strain field gradient during the plastic deformation process, and the back stress was generated from the strain field gradient. Fig. 15 (a) illustrated the fine grains and coarse grains in 40H characterized by BF-TEM. Lots of the dislocation sources in the fine grain zones facilitate the dislocations nucleation, and promote the activation of the slip system into a "soft phase" with high deformation ability under external forces [28]. As shown in Fig. 15 (b), fine grain zones possessed plenty of dislocation cells, while there were almost no dislocations in the coarse grains. Moreover, the density of nano-oxides in fine grains were higher than that in coarse grains, which contributed to higher back stress in fine grains [29]. Therefore, the activation from slip systems was further restrained by the back stress hardening in fine grains. Meanwhile, the flow stress was also activated by the back stress. As a result, the "soft phase" would be continuously hardened through the back stress hardening, which was beneficial to the increase of the strength. Nevertheless, the nucleation of dislocation in coarse grains was difficult due to the few dislocation sources, so the "hard phase" with the low deformability were prone to be formed. During the plastic deformation process, the back stress softening in "hard phase" would compensate the back stress hardening in "soft phase" to adapting to the strain fields incompatibility [30]. Therefore, slip systems in coarse grains were activated by the back stress softening, which contributed to the dislocation slipping from fine grains to the coarse grains and increased the plasticity. Moreover, Lee, et, al have found that these microcracks in bimodal alloys were inclined to be generated in fine grains because of the stress concentration, while coarse grains in bimodal alloys were essential to be the obstacles to prevent the crack propagation [31]. Thus, the ductility of ODS FeCrAl alloys in this work were significantly improved.

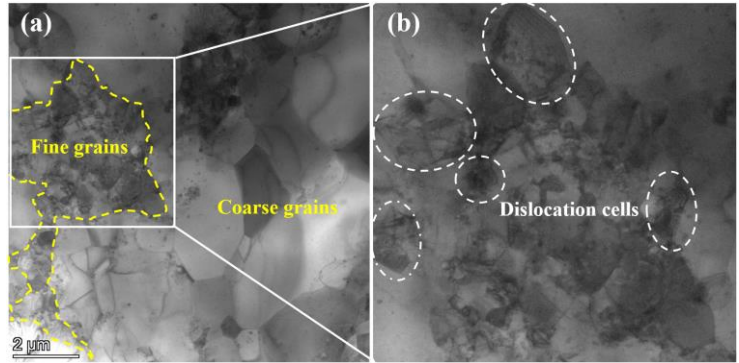

**Fig. 15**. (a) Fine grains and coarse grains in 40H, (b) Dislocation cells in fine grains.

### 4.4. Strengthening mechanism

Based on the microstructural characterization and mechanical properties tests, it was significant interest in this work to know how the fabrication technique and microstructure affect the strength of ODS FeCrAl alloys. The main strengthening mechanisms of ODS steels were expressed by five strengthening components: Peierls-Nabarro stress ($\sigma_o$), solid solution strengthening ($\sigma_{ss}$), grain boundary strengthening ($\sigma_{gb}$), dislocation strengthening ($\sigma_{dis}$), and nano-oxides strengthening ($\sigma_{particle}$) [32].

#### 4.4.1. Peierls-Nabarro stress

Peierls-Nabaro stress is the stress that causes dislocations to pass through a perfect lattice [33]. It is generally represented as

$$\sigma_0 = \frac{2M\mu}{1-v} \exp\left[\frac{-2\pi a}{b(1-v)}\right] \quad (1)$$

In the Peierls-Nabarro component of the matrix hardening, $v=0.3$ is the Poisson's ratio. $M=3$ is the Taylor factor. $b=0.248$ nm is the Burgers vector of the gliding dislocation system [34]. $a = 0.287$ nm is the lattice parameter [35]. $\mu=81$GPa is the shear modulus [36]. According to Eq. (1), $\sigma_o$ was calculated to be 55MPa.

#### 4.4.2. Solid-solution strengthening

The contribution of solid solution could be illustrated by

$$\sigma_{ss} = \sum k_i \cdot c_i^z \quad (2)$$

The solid solution component from the matrix hardening includes the sum of the constants $k_i$ with the concentration in every element '$i$' to the $z$ power, and $z=0.75$ [34]. Where $k_i$ is hardening constant, $c_i$ is the atom percent of substitutional elements. With

$k_{Cr}$=9.65MPa and $k_{Al}$=27.56MPa, and equivalent atomic concentrations of Cr and Al are 11.04% and 11.34%, respectively [34]. The strengthening component from Cr and Al in this study is identified as 229MPa for ODS FeCrAl alloys.

### 4.4.3. Grain boundary strengthening

The contribution from grain boundaries to alloy strength is usually represented by the Hall Petch relationship, which is expressed by:

$$\sigma_{gb} = \frac{k_{HP}}{\sqrt{D}} \tag{3}$$

where $k_{HP}$ = 401MPa.m$^{-1/2}$ is the Hall-Petch coefficient for ODS FeCrAl alloys [10]. The average grain diameters of $D$ are 15 μm, 14.2 μm, and 10.3 μm for 10H, 20H, and 40H, respectively. The strengthening values from grain boundary can be calculated to be 79MPa, 81MPa, and 95MPa for 10H, 20H, and 40H, respectively.

### 4.4.4. Dislocation strengthening

The dislocation hardening can be depicted as follows:

$$\sigma_{dis} = M\alpha\mu b\sqrt{\rho_{dis}} \tag{4}$$

where $b$, $M$, and $\mu$ are the same as listed previously and $α$=0.38 is a constant for the ferritic alloy [36]. However, dislocation density varied greatly with the processing route. In addition, Bentley pointed out that dislocation distribution varied vastly among different grains, and some of the grains are even free of dislocation, which is coincidence to the results from our study in Fig. 15 [37]. Therefore, it is difficult to be accurate calculation. Literature about various ODS steels (14YWT, MA957, 9Cr) have found out the dislocation densities varied between 1.6-2.4×10$^{14}$/m$^2$ by in-situ X-ray diffraction experiments at room temperatures [38]. However, the grain sizes in this work were larger than those in the literature, so the pinning effect of the dislocation was weaker and the dislocation density was smaller. In addition, Ren found that the density was 5.8×10$^{13}$/m$^2$ at the consolidation temperature of 1150 °C, and the dislocation density was expected to increase in a lower consolidation temperature [33]. Therefore, it made sense that the dislocation density of 10$^{14}$/m$^2$ was appropriate and reasonable for ODS FeCrAl alloys in this work. The dislocation strengthening was calculated to be 237MPa in this work. Although this assumption in this study will undoubtedly overestimate or underestimate the yield strength affected by the dislocation strengthening, it is still possible to analyze the general trends of the current system in comparing with the experimental results.

### 4.4.5. Nano-oxides strengthening

Deriving the expression for the nano-oxides strengthening required significant assumptions since the interaction between nano-oxide and dislocations was not yet clear. In generally, there existed two kinds of the strengthening mechanism from nano-oxides. One is the Orowan by-pass mechanism [39-41], and the other is the dispersed barrier hardening [42].

Orowan by-pass mechanism in ODS steels resumed that nano-oxides were impenetrable incoherent particles, and the nano-oxides strengthening from the Orowan dislocation by-pass mechanism could be followed by

$$\sigma_{particle} = \frac{0.176 M \mu b f^{1/2}}{d} ln \frac{d}{2b} \tag{5}$$

where $b$, $M$, are $\mu$ are the same as listed previously, $f$ and $d$ are the volume fraction and the average diameter of nano-oxides, respectively [43].

Dispersed barrier hardening is based on the direct geometric considerations of barrier layers intersecting with dislocation slip planes, which is described by

$$\sigma_{particle} = M\alpha(r)\mu b\sqrt{2Nr_p} \tag{6}$$

$$\alpha(r) = -0.017 + 0.374 log_{10} \frac{r_p}{2b} \tag{7}$$

where $b$, $M$, are $\mu$ are the same as listed previously, $\alpha(r)$ is the barrier strength coefficient, which represents the strength from the defect clusters to prevent dislocation motion [10]. It was reported that the small nano-oxides sheared by dislocations were relatively soft obstacles when the value of $\alpha(r)$ was between approximately 0.05 and 0.3 [3, 36].

In brief, the equations for strengthening by 'hard' and 'soft' particles are presented. 'Hard' particles from Orowan equation describes the nano-oxides strengthening due to 'hard' particles being by passed by dislocations, and the 'soft' particles become deformed when plastic yielding occurs. Hence, the value of α(r) is an important parameter for determining 'soft' or 'hard' particles [43]. In this work, α(r) of the barrier strength coefficient were 0.52 and 0.40 in 10H and 20H, respectively, so Orowan by-pass mechanism was appropriated to obtain the nano-oxides strengthening. However, the α(r) was 0.27 in 40H, which was coincidence with the soft particles from dispersed barrier hardening. Therefore, the values from nano-oxides strengthening were 236MPa, 336MPa, and 399MPa in 10H, 20H, and 40H, respectively.

Many scholars have studied the strengthening mechanism from nano-oxides in ODS steels. Zhou, et, al investigated the contribution of Zr content to the improvement of the mechanical properties in ODS FeCrAl alloys. Orowan by-pass mechanism were described to obtain the nano-oxide strengthening between 100MPa and 400MPa in (0-1.2Zr) ODS FeCrAl alloys. In fact, the α(r) were from 0.32 to 0.79 in (0-1.2Zr) ODS FeCrAl alloys, which was the hard particle [12]. And also, Li surveyed the influence of hot rolling process on mechanical properties of 15Cr-4Al ODS FeCrAl alloys, and Orowan by-pass mechanism were also employed to obtain the nano-oxides strengthening. The contribution of nano-oxides strengthening to yield strength were 612MPa and 642MPa, when the particle diameters were 12.6 nm and 12 nm with the density of $3.6 \times 10^{22}/m^3$ and $3.6 \times 10^{22}/m^3$ in ODS FeCrAl alloys before and after hot rolling, respectively [11]. And we found the α(r) was 0.39, which was beyond the scope of soft particles in Eq. (7). In addition, Massey, et, al. fabricated ODS FeCrAl alloys with high density ($1.8 \times 10^{23}/m^3$) and fine size (3.9 nm) of nano-oxides by MA and HE processes. However, the coefficient of the barrier strength was 0.21, which represented the soft particles that deviate significantly from fully impenetrable Orowan obstacles [10]. To sum up, it is appropriate and valid to judge soft or hard particles by the range of the barrier strength coefficient.

The contribution from various strengthening mechanisms could be expressed in Tab. 3. There were two types of superposition methodologies to obtain the total strengthening for different ODS steel [32, 33, 44], as shown in Eq. (8) and Eq. (9).

**Tab. 3.** Estimated various kinds of strength for ODS FeCrAl alloys (MPa).

| Samples | $\sigma_0$ | $\sigma_{ss}$ | $\sigma_{gb}$ | $\sigma_{dis}$ | $\sigma_{particle}$ |
|---------|------------|---------------|---------------|----------------|---------------------|
| **10H** | 55 | 229 | 79 | 237 | 236 |
| **20H** | 55 | 229 | 81 | 237 | 336 |
| **40H** | 55 | 229 | 95 | 237 | 399 |

$$\sigma_{YS} = \sigma_0 + \sigma_{ss} + \sigma_{gb} + \sigma_{dis} + \sigma_{particle} \tag{8}$$

$$\sigma_{YS} = \sigma_0 + \sigma_{ss} + \sigma_{gb} + \sqrt{\sigma_{dis}^2 + \sigma_{particle}^2} \tag{9}$$

The yield strength from the different superposition methodologies and experiment values were plotted in Fig. 16. The estimated value from Eq. (8) were obviously higher than that of the experimental yield strength, which meant that linear theoretical hardening models in Eq. (8) was not suitable to predict the strengthening mechanisms in this work. Nevertheless, the value in Eq. (9) matched well with the experimental value, indicating that it was effective to evaluate the yield strength. In fact, it conveyed that the strengthening mechanisms from nano-oxides was interrelated with the dislocation strengthening. As we all know, nano-oxides could prevent the moving of dislocation, and further affect the mean free path in dislocation, which were observed in Fig. 8(c) in this work. Hence, the two mechanisms were mutually restricted. Furthermore, we observed that the difference between the estimated (Eq. 9) and experimental values gradually increased, as the milling time increases. On the one hand, a constant of dislocation density of $10^{14}/m^2$ in this study were employed in three kinds of ODS FeCrAl alloys, while the dislocation density would gradually increase due to more input energy from a longer milling time. In addition, more and smaller nano-oxides and nano-clusters would precipitate as the ball milling time increased, which was difficult to be accurately observed and counted through TEM characterization. More importantly, small nano-clusters were more likely to maintain a coherent orientation relationship with the matrix, which were contribute to the yield strength through the coherent strengthening mechanism [8, 45]. Therefore, we will attempt to obtain more accurate nano-clusters information through more advanced characterization methods, which can more accurately predict the strengthening mechanism of ODS FeCrAl alloys in the future.

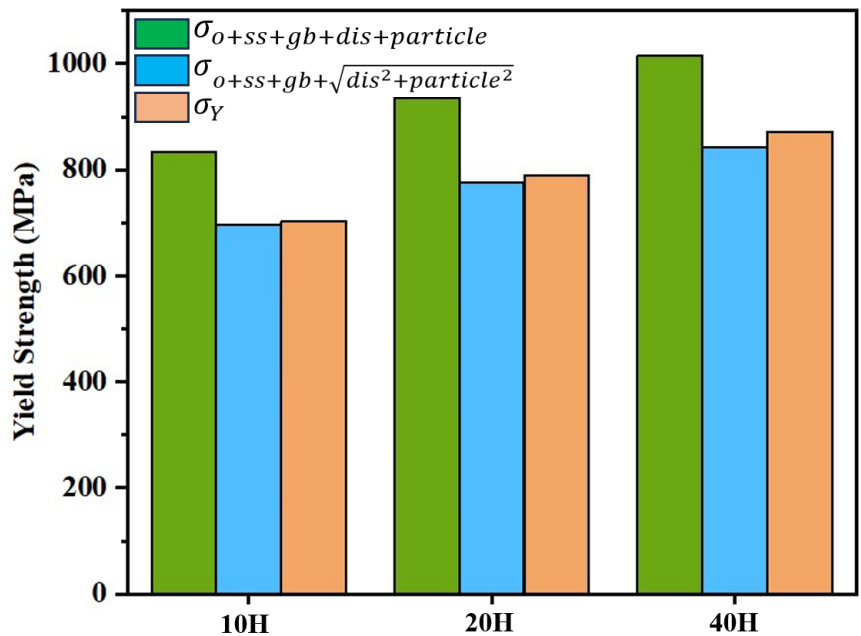

**Fig. 16.** Experiment measured yield strength and the theoretically calculated values.

## 5. Conclusion

ODS FeCrAl alloys with various milling time (10h, 20h, and 40h) were prepared by MA and SPS in this study. The main conclusions are summarized as follows:

(1) The bimodal grain distribution was introduced into the ODS FeCrAl alloys by the consolidated by SPS, and average grain sizes decreased with the increase of milling time. The grain sizes were 15μm, 14.3μm, and 10.2μm for 10H, 20H, and 40H, respectively.

(2) The size of nano-oxides were 27.2nm, 13.1nm and 5.9nm and the density were 1.44 ×$10^{22}/m^3$, 5.70×$10^{22}/m^3$, 9.61×$10^{22}/m^3$ in the samples of 10H, 20H, and 40H, respectively. Moreover, large-size nano-oxides were $Y_3Al_5O_{12}$, while the fine nano-oxides were prone to be $YTiO_3$, $Y_2Zr_2O_7$, and $Y_4Zr_3O_{12}$.

(3) The mechanical and microstructural properties of these ODS FeCrAl alloys were highly dependent on the milling time. And the mechanical properties test showed that the strength and total elongation gradually increase with increasing milling time. Ultimate strength was 993MPa with the total elongation of 27.65% in 40H, which showed exceptional strength with an excellence plasticity.

(4) The bimodal grains distributions play a key role in preventing the initiation and further propagation for these microcracks, and the effect of back stress was as well stimulated during plastic deformation process. Therefore, the synergistic improvement of both plasticity and strength were realized in this study.

(5) The nano-oxide strengthening in these samples were originated from the different ways. Orowan by-pass mechanism was appropriated in the samples of 10H and 20H, while the sample of 40H was suitable for the dispersed barrier hardening. In addition, the strengthening mechanism derived from nano-oxides strengthening were interrelated with the dislocation strengthening, which was more appropriate to predict reinforcement mechanism models than that in simple linear superposition. Above all, the SPS preparation process by bimodal grain distribution is an effective approach to both improve strength and plasticity in ODS FeCrAl alloys.

**Acknowledgment**

This study is finacially supported by Innovative Scientific Program of CNNC; the National Natural Science Foundation of China under Grant No. 12375159; GuangDong Basic and Applied Basic Research Foundation under Grant No. 2020B1515120035.